\documentclass[english]{aa}
\usepackage[utf8]{inputenc}
\usepackage{txfonts}
\usepackage{booktabs}
\usepackage[usenames]{color}
\usepackage{graphicx}
\usepackage{amssymb}
\IfFileExists{url.sty}{\usepackage{url}}
                      {\newcommand{\url}{\texttt}}
\usepackage[authoryear]{natbib}

\makeatletter
\providecommand{\tabularnewline}{\\}

\usepackage[breaklinks, colorlinks]{hyperref}
\hypersetup{
pdfauthor = {Konrad R. W. Tristram},
pdftitle  = {On the size-luminosity relation of AGN dust tori in the mid-infrared},
citecolor = blue,
linkcolor = blue,
urlcolor  = Sepia
}
\usepackage[all]{hypcap}

\let\HyperRaiseLinkLength\@tempdima
\let\HyperRaiseLinkHook\@empty
\renewcommand{\HyperRaiseLinkDefault}{1.5\baselineskip}
\def\Hy@raisedlink#1{%
\setlength\HyperRaiseLinkLength\HyperRaiseLinkDefault
\HyperRaiseLinkHook
\ifvmode
#1%
\else
\penalty\@M
\smash{\raise\HyperRaiseLinkLength\hbox{#1}}%
\fi
}

\bibpunct{(}{)}{;}{a}{}{,}

\makeatother

\begin{document}

\title{On the size--luminosity relation of AGN dust tori in the mid-infrared\thanks{Based on observations with MIDI at the European Southern Observatory, Chile, programme numbers 060.A-9224(A), 074.B-0213(B), 075.B-0697(B), 076.B-0038(A), 076.B-0743(A,C), 077.B-0026(B), 078.B-0031(A), 079.B-0180(A), 080.B-0258(A), 081.D-0092(A) and 381.B-0240(A,B).}}

\author{K. R. W. Tristram\inst{1} and M. Schartmann\inst{2,3}}

\institute{Max-Planck-Institut für Radioastronomie, Auf dem Hügel 69, 53121
Bonn, Germany \\
e-mail: \href{mailto:tristram@mpifr-bonn.mpg.de}{\nolinkurl{tristram@mpifr-bonn.mpg.de}}\and Max-Planck-Institut
für extraterrestrische Physik, Giessenbachstraße, D-85748 Garching,
Germany\\
e-mail: \href{mailto:schartmann@mpe.mpg.de}{\nolinkurl{schartmann@mpe.mpg.de}}\and Universitäts-Sternwarte
München, Scheinerstraße 1, D-81679 München, Germany}

\titlerunning{On the size--luminosity relation of AGN dust tori}

\authorrunning{K. R. W. Tristram \& M. Schartmann}

\abstract{Interferometric measurements in the mid-infrared have shown that
the sizes of the warm dust distributions in active galactic nuclei
are consistent with their scaling with the square root of their luminosity.} {
We carry out a more detailed analysis of this size--luminosity relation
to investigate which of the general properties of the dusty tori in
active galactic nuclei can be derived from this relation. We are especially
interested in the cases, where only a very small number of interferometric
measurements are available and the sizes are directly calculated from
the measured visibilities assuming a Gaussian brightness distribution.} {
We improve the accuracy of the size--luminosity relation by adding
a few additional size measurements from more recent interferometric
observations and compare the measured sizes to those derived from
hydrodynamical and radiative transfer models of AGN tori.} {We find
that a Gaussian approximation yields a reasonable estimate of the
size of the brightness distribution, as long as the visibilities are
within $0.2\leq V\leq0.9$. The uncertainty in the size estimate depends
on the true brightness distribution and is up to a factor of four
for the models used in our investigation. The size estimates derived
from the models are consistent with those determined from the measurements.
However, the models predict a significant offset between the sizes
derived for face-on (Seyfert~1 case) and edge-on (Seyfert~2 case)
tori: the face-on tori should appear significantly more compact for
the same luminosity. This offset is not observed in the current data,
probably because of the large uncertainties and low statistics of
the present interferometric measurements. Furthermore, we find a ratio
of the mid- to near-infrared sizes of approximately 30, whereas the
first probes the body of the torus and the second is an estimate of
the inner rim.} { The size--luminosity relation of AGN tori using
Gaussian size estimates is a very simple and effective tool to investigate
the internal structure and geometry of AGN tori and obtain constraints
on the differences between type~1 and type~2 AGN. However, to fully
exploit the possibilities of investigating the nuclear distributions
of gas and dust in AGN using this size--luminosity relation, more
accurate interferometric measurements of a larger sample of AGN are
needed.}


\date{Received March 11, 2011; accepted April 28, 2011}

\maketitle

\section{Introduction\label{sec:introduction}}

A toroidal distribution of warm gas and dust is a key component of
active galactic nuclei (AGN). In unified schemes of AGN \citep[e.g.][]{1993Antonucci},
this so-called \emph{dusty torus} provides the viewing angle dependent
obscuration of the central engine of AGN as well as the material for
feeding the supermassive black hole. The dusty torus absorbs the hard
continuum emission from the accretion disk and reemits it in the infrared.
This direct connection between the emission from the accretion disk
and the reprocessed emission from the torus is reflected by the tight
correlation between the X-ray and the mid-infrared luminosities for
both Seyfert 1 and Seyfert 2 galaxies \citep[e.g.][]{2008Horst1,2009Gandhi,2009Levenson}.

It was realised quite early that the nuclear dust is most likely distributed
in clumps \citep{1988Krolik}. However, because of a lack of computational
power as well as of suitable radiative transfer codes, a first attempt
to model clumpy dust distributions in AGN was only undertaken by \citeauthor{2002Nenkova}
in 2002. Subsequent modelling of clumpy tori was carried out by \citet{2006Hoenig},
\citet{2008Nenkova1,2008Nenkova2}, \citet{2008Schartmann,2009Schartmann,2010Schartmann},
and \citet{2010Hoenig2}.%
\begin{table*}
\caption{Galaxy properties and characteristics derived from the interferometric
observations of the AGN studied by MIDI.\label{table:source_list}}

\centering

\begin{tabular}{@{ \extracolsep{1.0pt}}l@{ \extracolsep{3.0pt}}l@{ \extracolsep{3.0pt}}r@{ \extracolsep{3.0pt}}r@{ \extracolsep{3.0pt}}r@{ \extracolsep{3.0pt}}r@{ \extracolsep{3.0pt}}r@{ \extracolsep{3.0pt}}r@{ \extracolsep{3.0pt}}r}
\toprule 
\hline Galaxy &  & $D$ & $L_{\mathrm{X}}$ & modified  & $L_{\mathrm{MIR}}$ & $\mathit{BL}$ & $V$ & $s_{12\mathrm{\mu m}}$\tabularnewline
name & type & ~{[}Mpc] & {[}W] & Julian Date & {[}W] & {[}m] &  & {[}pc]\tabularnewline
(1) & (2) & (3) & (4) & (5) & (6) & (7) & (8) & (9)\tabularnewline
\midrule
\object{NGC~1068} & Sy~2 & 14 & ~~~~$2.1\cdot10^{36}$ & ~~~~$53687.04006$ & ~~~~$(1.0\pm0.1)\cdot10^{37}$ & ~~~~$79.9$ & ~~~~$0.07\pm0.03$ & $>1.8$\tabularnewline
\object{NGC~1068}\vspace{1mm}
 & Sy~2 & 14 & $2.1\cdot10^{36}$ & $53688.02715$ & $(1.0\pm0.1)\cdot10^{37}$ & $33.6$ & $0.24\pm0.06$ & $3.2$\tabularnewline
\object{NGC~1365} & Sy~1.8 & 18 & $5.7\cdot10^{35}$ & $53989.32147$ & $(5.1\pm2.2)\cdot10^{35}$ & $46.6$ & $0.51\pm0.10$ & $2.0$\tabularnewline
\object{NGC~1365} & Sy~1.8 & 18 & $5.7\cdot10^{35}$ & $54428.07902$ & $(3.4\pm3.1)\cdot10^{35}$ & $54.4$ & $0.78\pm0.57$ & $<2.7$\tabularnewline
\object{NGC~1365}\vspace{1mm}
 & Sy~1.8 & 18 & $5.7\cdot10^{35}$ & $54428.16083$ & $(4.7\pm4.3)\cdot10^{35}$ & $62.3$ & $0.71\pm0.21$ & $1.1$\tabularnewline
\object{MCG-05-23-016} & Sy~2 & 35 & $3.3\cdot10^{36}$ & $53723.27767$ & $(2.2\pm0.9)\cdot10^{36}$ & $46.1$ & $0.70\pm0.19$ & $2.8$\tabularnewline
\object{MCG-05-23-016}{*}\vspace{1mm}
 & Sy~2 & 35 & $3.3\cdot10^{36}$ & $54575.03573$ & $(3.6\pm3.3)\cdot10^{36}$ & $129.6$ & $0.26\pm0.09$ & $>1.8$\tabularnewline
\object{Mrk~1239} & Sy~1.5 & 80 & $3.5\cdot10^{37}$ & $53723.31388$ & $(1.1\pm0.4)\cdot10^{37}$ & $43.0$ & $0.88\pm0.17$ & $<7.0$\tabularnewline
\object{Mrk~1239}{*} & Sy~1.5 & 80 & $3.5\cdot10^{37}$ & $54576.02468$ & $(1.2\pm0.8)\cdot10^{37}$ & $127.4$ & $0.60\pm0.15$ & $2.9$\tabularnewline
\object{Mrk~1239}{*}\vspace{1mm}
 & Sy~1.5 & 80 & $3.5\cdot10^{37}$ & $54577.06465$ & $(1.5\pm0.5)\cdot10^{37}$ & $100.7$ & $0.68\pm0.15$ & $3.1$\tabularnewline
\object{NGC~3783} & Sy~1 & 40 & $4.1\cdot10^{36}$ & $53521.12588$ & $(6.0\pm0.8)\cdot10^{36}$ & $68.6$ & $0.40\pm0.15$ & $3.6$\tabularnewline
\object{NGC~3783}\vspace{1mm}
 & Sy~1 & 40 & $4.1\cdot10^{36}$ & $53521.14027$ & $(4.9\pm0.7)\cdot10^{36}$ & $64.9$ & $0.54\pm0.15$ & $3.1$\tabularnewline
\object{NGC~4151} & Sy~1.5 & 14 & $1.5\cdot10^{36}$ & $54138.33590$ & $(6.7\pm5.8)\cdot10^{35}$ & $35.8$ & $0.22\pm0.10$ & $>2.6$\tabularnewline
\object{NGC~4151}{*} & Sy~1.5 & 14 & $1.5\cdot10^{36}$ & $54578.09375$ & $(8.3\pm2.1)\cdot10^{35}$ & $61.1$ & $0.29\pm0.06$ & $1.6$\tabularnewline
\object{NGC~4151}{*}\vspace{1mm}
 & Sy~1.5 & 14 & $1.5\cdot10^{36}$ & $54580.12176$ & $(9.2\pm5.6)\cdot10^{35}$ & $89.1$ & $0.21\pm0.07$ & $>1.1$\tabularnewline
\object{3C~273} & QSO & 650 & $2.6\cdot10^{39}$ & $54138.25289$ & $(4.0\pm1.3)\cdot10^{38}$ & $36.7$ & $0.90\pm0.20$ & \textcolor{white}{${\normalcolor <67.}{\color{white}3}$}\tabularnewline
\object{3C~273}\vspace{1mm}
 & QSO & 650 & $2.6\cdot10^{39}$ & $54578.28634$ & $(3.6\pm3.2)\cdot10^{38}$ & $30.7$ & $0.74\pm0.22$ & ~~~\textcolor{white}{${\normalcolor <110.}{\color{white}4}$}\tabularnewline
\object{IC~4329A} & Sy~1.2 & 65 & $1.9\cdot10^{37}$ & $54138.36383$ & $(1.3\pm1.1)\cdot10^{37}$ & $46.6$ & $0.76\pm0.53$ & $<10.8$\tabularnewline
\object{IC~4329A}{*} & Sy~1.2 & 65 & $1.9\cdot10^{37}$ & $54576.11638$ & $(1.4\pm0.5)\cdot10^{37}$ & $125.3$ & $0.51\pm0.08$ & $2.7$\tabularnewline
\object{IC~4329A}{*} & Sy~1.2 & 65 & $1.9\cdot10^{37}$ & $54577.11216$ & $(1.5\pm0.6)\cdot10^{37}$ & $102.0$ & $0.72\pm0.17$ & $2.3$\tabularnewline
\object{IC~4329A}{*}\vspace{1mm}
 & Sy~1.2 & 65 & $1.9\cdot10^{37}$ & $54577.27403$ & $(1.0\pm0.3)\cdot10^{37}$ & $95.4$ & $0.66\pm0.09$ & $2.8$\tabularnewline
\object{Circinus galaxy} & Sy~2 & 4 & $4.6\cdot10^{35}$ & $53159.33199$ & $(4.9\pm0.4)\cdot10^{35}$ & $20.7$ & $0.18\pm0.04$ & $>1.5$\tabularnewline
\object{Circinus galaxy}\vspace{1mm}
 & Sy~2 & 4 & $4.6\cdot10^{35}$ & $53430.42248$ & $(4.9\pm0.4)\cdot10^{35}$ & $62.4$ & $0.11\pm0.02$ & $>0.6$\tabularnewline
\object{NGC~7469} & Sy~1.2 & 65 & $4.0\cdot10^{36}$ & $53989.25648$ & $(8.8\pm7.7)\cdot10^{36}$ & $46.4$ & $0.25\pm0.06$ & $10.5$\tabularnewline
\bottomrule
\end{tabular}

~

\begin{minipage}[t][1\totalheight]{1\textwidth}%
Notes: The columns are (1) the name, (2) the type and (3) the distance
$D$ of the galaxy; (4) the estimated intrinsic AGN luminosity $L_{X}$;
(5) the modified Julian Date ($JD-2400000.5$) of the interferometric
observation; (6) the mid-infrared luminosity $L_{\mathrm{MIR}}$ derived
from the mid-infrared flux measured by MIDI; (7) the projected baseline
length, $\mathit{BL}$, for the respective interferometric observation;
(8) the visibility $V$ at $12\,\mu\mathrm{m}$; and (9) the approximate
size $s_{12\mathrm{\mu m}}$ of the emitter at $12\,\mu\mathrm{m}$.
For $V\lesssim0.2$ and $V\gtrsim0.9$ (including the uncertainties
in $V$), only upper or lower limits to the size can be given. The
new data points -- in addition to those already published in \citet{2009Tristram}
-- are marked by an asterisk ({*}) after the name.%
\end{minipage}%

\end{table*}

Observationally, interferometry in the near- and the mid-infrared
led to a breakthrough in the investigation of dusty tori by being
able to resolve the nuclear infrared emission of the warm dust. Using
interferometric methods, it is now possible to directly study the
physical properties of the nuclear dust distributions in several AGN
and compare models of the torus with spatial information on its infrared
emission. Interferometric studies of individual sources have been
published for \object{NGC~4151} \citep{2003Swain,2009Burtscher,2010Pott},
\object{NGC~1068} \citep{2004Wittkowski,2004Jaffe1,2006Poncelet,2009Raban},
\object{Centaurus~A} \citep{2007Meisenheimer,2010Burtscher}, the
\object{Circinus galaxy} \citep{2007Tristram2}, and \object{NGC~3783}
\citep{2008Beckert}. First attempts have been undertaken to study
several sources in a more generic way and to derive the general properties
of the dust distributions. In \citet{2009Kishimoto1}, a possible
common radial structure for the dust distribution in Seyfert~1 nuclei
was proposed, while in \citep{2009Kishimoto2,2011Kishimoto}, the
hot inner rim of the dust distribution was probed for several type
1 AGN (among them \object{NGC~4151}). In \citet{2009Tristram},
mid-infrared interferometry of several type 1 and type 2 AGN is presented
and the discovery of a size--luminosity relation has been claimed
for the dust distributions in the mid-infrared: at $12\,\mu\mathrm{m}$,
the size $s$ of the torus was found to be consistent to scale with
the monochromatic luminosity $L$ as $s=p\cdot L^{0.5}$ and the proportionality
factor of the relation was determined as $p=\left(1.8\pm0.3\right)\cdot10^{-18}\,\mathrm{pc}\cdot\mathrm{W}^{-0.5}$.

This paper presents a follow-up study of this size--luminosity relation
by including new data and by comparison of the measured relation to
the predictions of hydrodynamic and radiative transfer models of AGN
tori. The aim is to investigate the degree to which simple size estimates
assuming Gaussian brightness distributions can be used to constrain
torus models and investigate the torus structure.

The paper is organised as follows. In Sect.~\ref{sec:data} the data
used to determine the sizes of the dust distributions and the AGN
luminosity are described. In Sect.~\ref{sec:modelling}, the models
used for a comparison to the data are presented. The results and their
discussion are found in Sect.~\ref{sec:discussion} and our conclusions
are given in Sect.~\ref{sec:conclusions}.

\section{Data\label{sec:data}}

The data used to determine the size--luminosity relation was obtained
with the MID-infrared Interferometric instrument (MIDI) at the Very
Large Telescope Interferometer (VLTI) using the $8.2\,\mathrm{m}$
Unit Telescopes (UTs) of the array. Most of these were previously
presented in \citet{2009Tristram}. Additional data from more recent
MIDI observations of \object{NGC~4151}, \object{Mrk~1239}, \object{MCG-05-23-016},
and \object{IC~4329A} were included in the present analysis to
obtain a larger statistical sample of size estimates for these five
sources. The new data was processed in exactly the same way as the
original data and we refer to \citet{2009Tristram} for a description
of the data reduction procedures and settings. Only for the second
measurement of \object{3C~273} on 2008 Apr 22 ($mJD=54578.286$)
was the reduction carried out with a slightly modified mask. However,
this leads only to an insignificant change in the resulting fluxes
of this source. A detailed analysis and modelling of the interferometric
data for \object{NGC~4151} can be found in \citet{2009Burtscher}.
A full presentation and analysis of the new data for \object{MCG-05-23-016},
\object{Mrk~1239}, and \object{IC~4329A} will be presented elsewhere.

As in \citet{2009Tristram}, the sizes of the dust emitters are estimated
by assuming a Gaussian brightness distribution. Although the true
brightness distributions of AGN tori are significantly more complex
than a Gaussian brightness distribution, a Gaussian is the simplest
and most general initial guess for the true brightness distribution
and allows a very straightforward determination of the spatial extent
of the source. The full width at half maximum $\mathit{FWHM}(\lambda)$
of the Gaussian brightness distribution can be directly calculated
from the visibility $V(\lambda)$ and the projected baseline length
$\mathit{BL}$ of the interferometric observation according to\begin{equation}
\mathit{FWHM}(\lambda)=\frac{\lambda}{\mathit{BL}}\cdot\frac{2}{\pi}\sqrt{-\ln2\cdot\ln V(\lambda)},\label{eq:gauss-fwhm}\end{equation}
where $\lambda$ is the wavelength at which the calculation is carried
out and $\mathit{FWHM}$ is in radians. For the present investigation,
we follow \citet{2009Tristram} and adopt $\lambda=12\,\mathrm{\mu m}$.
At this wavelength, changes in the atmospheric transmission are the
smallest of all wavelengths measured by MIDI. At the same time the
size estimate is least affected by spectral features (i.e. by either
the silicate feature or line emission). As long as $0.2\lesssim V\lesssim0.9$,
the $\mathit{FWHM}$ derived in this way is a good estimate of the
emission region size. Otherwise only upper or lower limits to the
size can be given. The sizes derived by this simple method agree with
those determined from a more detailed modelling of the sources in
the cases of \object{NGC~1068} ($1\,\mathrm{pc}<s<4\,\mathrm{pc}$,
\citealt{2009Raban}), the \object{Circinus galaxy} ($0.4\,\mathrm{pc}<s<2\,\mathrm{pc}$,
\citealt{2007Tristram2}), and \object{NGC~4151} ($s=2.0\pm0.4\,\mathrm{pc}$,
\citealt{2009Burtscher}). A more detailed discussion of the accuracy
of this simple size estimate will be given in Sect.~\ref{subsec:modelled-sizes}.

The mid-infrared luminosities, $L_{\mathrm{MIR}}$, are directly calculated
from the mid-infrared fluxes measured with MIDI, and hence they have
very large errors. The accuracies of both the luminosities as well
as the visibilities (and hence also the size estimates) are limited
by the large uncertainties in the total fluxes measured with MIDI.
These uncertainties are caused by the insufficient removal of background
emission from the long optical train through the delay lines by the
chopping \citep[c.f.][]{2004Absil}. Owing to the non-linear form
of Eq.~\ref{eq:gauss-fwhm}, the errors in the size estimates are
calculated independently from the upper or lower limits to the visibility.

In addition to the mid-infrared data, we compile luminosities derived
from hard X-ray measurements for our sample of AGN. When corrected
for foreground and intrinsic absorption, the X-ray luminosity is considered
to be an effective proxy of the emission from the accretion disk,
that is, of the intrinsic AGN luminosity. To obtain a preferably
homogeneous set of values, most of the X-ray luminosities were taken
from the first 22 months of data of the hard X-ray survey ($14-195\,\mathrm{keV}$)
with the Burst Alert Telescope (BAT) on the Swift satellite \citep{2010Tueller}.
These measurements were obtained before or contemporaneous with the
interferometric measurements. We do not use the data for the 54-month
release of the hard X-ray survey with BAT \citep{2010Cusumano}, because
most of these data were obtained \textit{after} our interferometric
measurements. We nevertheless use the discrepancies between the 22-month
and 54-month catalogues of BAT data to estimate the general uncertainty
in the X-ray luminosities and find them to be of approximately $0.2\,\mathrm{dex}$.
This also takes into account a possible (moderate) variability of
the sources. Only one of our sources, \object{Mrk~1239}, is not
in the catalogue of sources from \citet{2010Tueller}, hence we use
the luminosity from \citet{2004Grupe}, $\log(L_{0.2-12.0\mathrm{keV}}/\mathrm{W})=37.29$
for this galaxy. Correction from the soft to the hard X-ray band using
the factor of $1.81$ from \citet{2009Rigby} with an additional correction
for the different soft X-ray bands assuming a flat power law finally
gives an estimate for the intrinsic luminosity of $L_{X}=3.5\cdot10^{37}\,\mathrm{W}$.

Three of our sources, \object{NGC~1068}, \object{NGC~1365},
and the \object{Circinus galaxy}, are significantly absorbed. Using
the hydrogen column densities $N_{\mathrm{H}}>1.0\cdot10^{25}\,\mathrm{cm}^{-2}$
\citep{2000Matt}, $N_{\mathrm{H}}=4.0\cdot10^{23}\,\mathrm{cm}^{-2}$
\citep{2005Risaliti1}, and $N_{\mathrm{H}}=4.3\cdot10^{24}\,\mathrm{cm}^{-2}$
\citep{2000Matt} as well as the ratios of emergent to input hard
X-ray fluxes from \citet{2009Rigby}, we obtain correction factors
of $>16$, $1.2$, and $3.6$ for the three galaxies, respectively.
These are applied to the measured luminosities to obtain the intrinsic
luminosities. For all other galaxies, the corrections due to extinction
along the line of sight are negligible.

All quantities (luminosities, baseline lengths, visibilities, and
size estimates) as well as the general properties of the galaxies
are summarised in Table~\ref{table:source_list}.

\section{Modelling\label{sec:modelling}}

To gain a clearer understanding of the implications of the interferometrically
measured sizes of the dust distributions, the size--luminosity relation
of dusty tori is compared to the predictions from models for clumpy
tori of AGN. We use two completely different modelling approaches
in the comparison to distinguish the features that are characteristic
of a certain model from more general properties of the dust distributions.
We interpret all results common to both models as an indication that
these results do not depend on the precise parameters of the respective
model but rather are intrinsic properties of a geometrically thick,
toroidal dust distribution.

\subsection{Full hydrodynamical torus model\label{subsec:hydro-models}}

\begin{figure*}
\centering

\includegraphics[bb=96bp 226bp 492bp 620bp,clip]{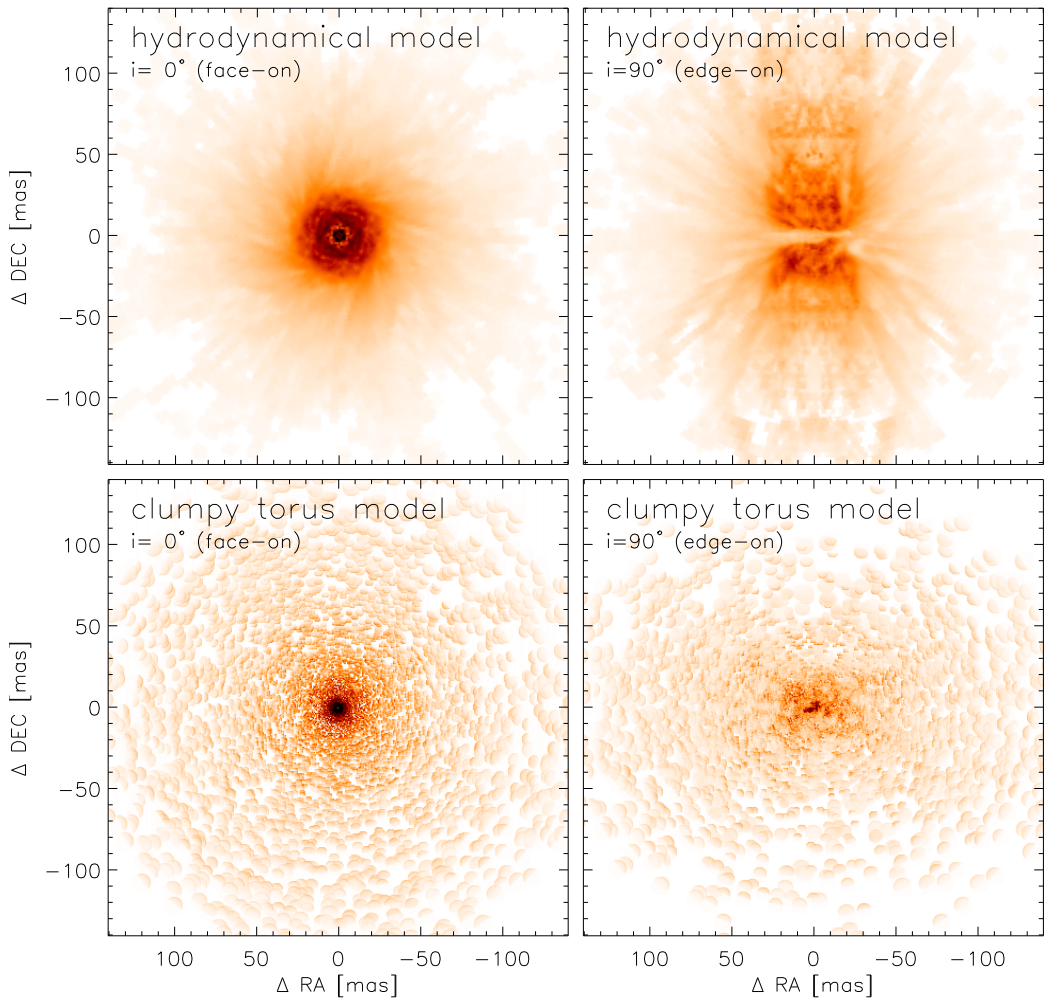}

\caption{Images of the torus models at $12\,\mathrm{\mu m}$ as used for our
size determinations. \textit{Top row}: full hydrodynamical torus model
(Sect.~\ref{subsec:hydro-models}); \textit{Bottom row}: clumpy torus
model (Sect.~\ref{subsec:rt-models}); \textit{Left panels}: Seyfert~1
($i=0^{\circ}$) case; \textit{Right panels}: Seyfert~2 ($i=90^{\circ}$)
case. For the example images shown here, a bolometric luminosity of
$L_{\mathrm{AD}}=4.6\cdot10^{37}\,\mathrm{W}$ and a distance of $D=45\,\mathrm{Mpc}$
were used for both models, in order to reach comparable sizes and
intensities. All images are plotted with the same intensity range
and a square root colour scale.}
\label{fig:model-images}
\end{figure*}

The first model is the full hydrodynamical simulation of dusty tori
in AGN from \citet{2009Schartmann}. This model accounts for both
the obscuration, as well as the feeding, of the central source. The
model follows the evolution of the interstellar medium by taking into
account discrete mass-loss and energy injection due to the evolution
of a nuclear star cluster, as well as optically thin radiative cooling.
The interplay between the injection of mass, supernova explosions,
and radiative cooling leads to a two-component structure consisting
of a cold geometrically thin, but optically thick and very turbulent
disk residing in the vicinity of the angular momentum barrier. The
disk is surrounded by a filamentary toroidal structure on scales of
tens of parsec. After the computationally very expensive hydrodynamical
simulations have been calculated, radiative transfer simulations are
carried out to obtain observable quantities such as surface brightness
distributions and spectral energy distributions (SEDs). For the present
comparison, we use the standard model from \citet{2009Schartmann}
with a bolometric luminosity for the accretion disk of $L_{\mathrm{AD}}=1.2\cdot10^{11}\, L_{\odot}=4.6\cdot10^{37}\,\mathrm{W}$,
which is meant to represent a typical Seyfert galaxy. Images of the
brightness distributions at $12\,\mathrm{\mu m}$ for the Seyfert~1
(face-on) and Seyfert~2 (edge-on) cases of this model are shown in
the top row of Fig.~\ref{fig:model-images}. For a more detailed
description of the hydrodynamical simulation as well as the subsequent
radiative transfer calculations, the reader is referred to \citet{2009Schartmann}.%

\subsection{Clumpy torus models\label{subsec:rt-models}}

The second torus model compared with the data is the model of \citet{2010Hoenig2},
which is an upgrade to the original 3D radiative transfer model of
clumpy dust tori presented in \citet{2006Hoenig}. The strategy of
this model is based on separating the simulations of individual cloud
SEDs and images, and the final SEDs and images of the torus. In a
first step, Monte Carlo radiative transfer simulations are used to
calculate the emission from individual clouds at different distances
from the centre. To finally simulate the entire torus emission, dust
clouds are randomly distributed around an AGN according to certain
physical and geometrical model parameters (for details see \citealt{2010Hoenig2}).
Here we use the model with the {}``spherical'' distribution of clouds
in the vertical direction. The six parameters used for the description
of the entire torus geometry are: (1) the radial dust-cloud distribution
power-law index $a$, (2) the half-opening angle $\theta_{0}$, (3)
the number of clouds along an equatorial line-of-sight $N_{0}$, (4)
the cloud radius at the sublimation radius $R_{\mathrm{cl}}$, (5)
the power-law index $b$ of the cloud sizes, and (6) the outer torus
radius $R_{\mathrm{out}}$. The inner torus radius, i.e., the sublimation
radius $R_{\mathrm{sub}}$, is defined by the bolometric luminosity
of the accretion disk, such that $R_{\mathrm{sub}}=1.1\cdot(L_{\mathrm{AD}}/10^{39}\,\mathrm{W})^{0.5}\,\mathrm{pc}$.
The parameter values used for the present model were derived from
the analysis of high resolution SEDs of Seyfert galaxies \citep{2010Hoenig1}
and are summarised in Table~\ref{tab:model_params}. Edge-on and
face-on images of this torus model, for the same luminosity as the
hydrodynamical model, are displayed in the bottom row of Fig.~\ref{fig:model-images}.
\begin{table}
\caption{Parameters of the clumpy torus model used for the present modelling.
See Sect.~\ref{subsec:rt-models} for a description of the parameters.\label{tab:model_params} }

\centering

\begin{tabular}{cc}
\hline 
\hline parameter & value\tabularnewline
\hline
$a$ & $-1$\tabularnewline
$\theta_{0}$ & $45^{\circ}$\tabularnewline
$N_{0}$ & $7$\tabularnewline
$R_{\mathrm{cl}}$ & $0.035\, R_{\mathrm{sub}}$\tabularnewline
$b$ & $1.0$\tabularnewline
$R_{\mathrm{out}}$ & $50\, R_{\mathrm{sub}}$\tabularnewline
\hline
\end{tabular}
\end{table}
\begin{figure*}
\centering

\includegraphics[bb=170bp 325bp 418bp 512bp]{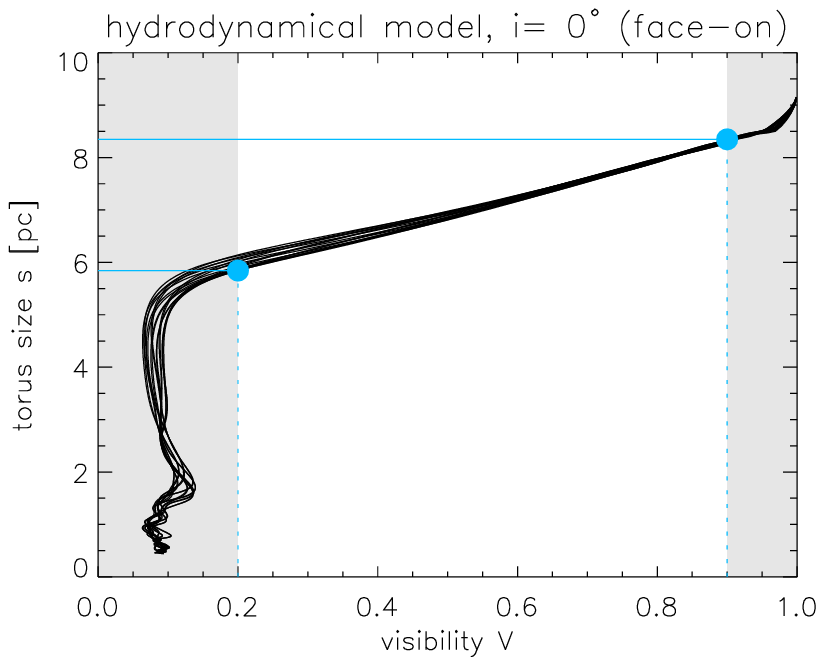}\hfill{}\includegraphics[bb=170bp 325bp 418bp 512bp]{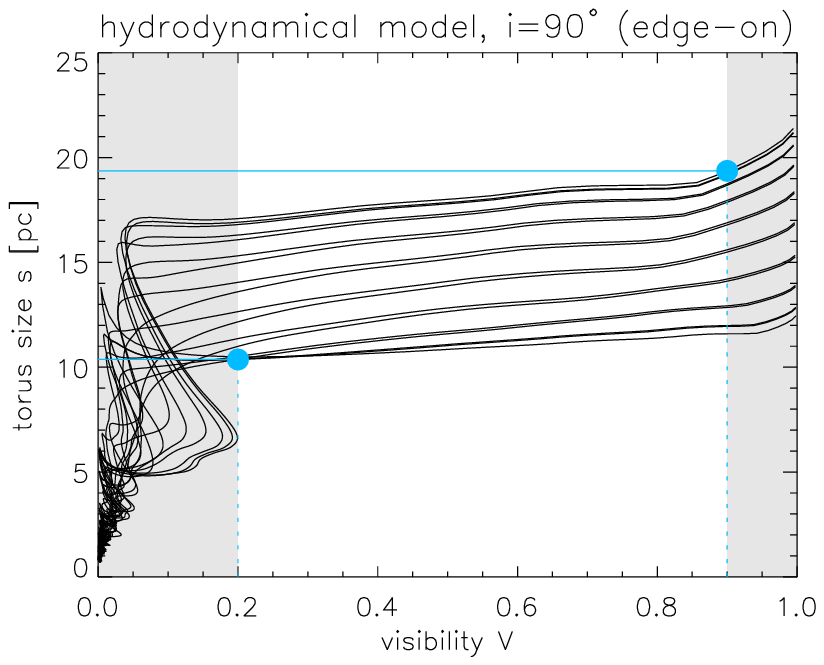}

~

\includegraphics[bb=170bp 325bp 418bp 512bp]{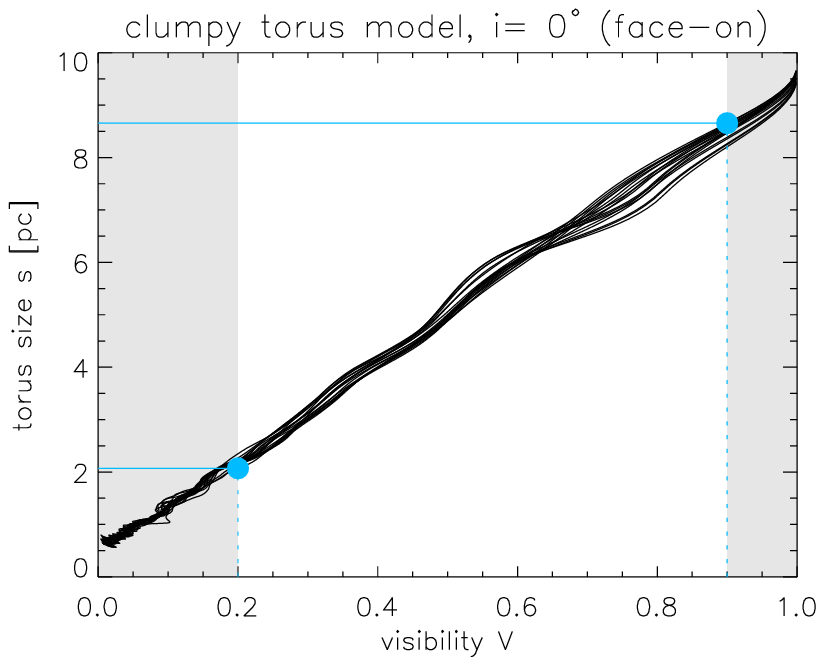}\hfill{}\includegraphics[bb=170bp 325bp 418bp 512bp]{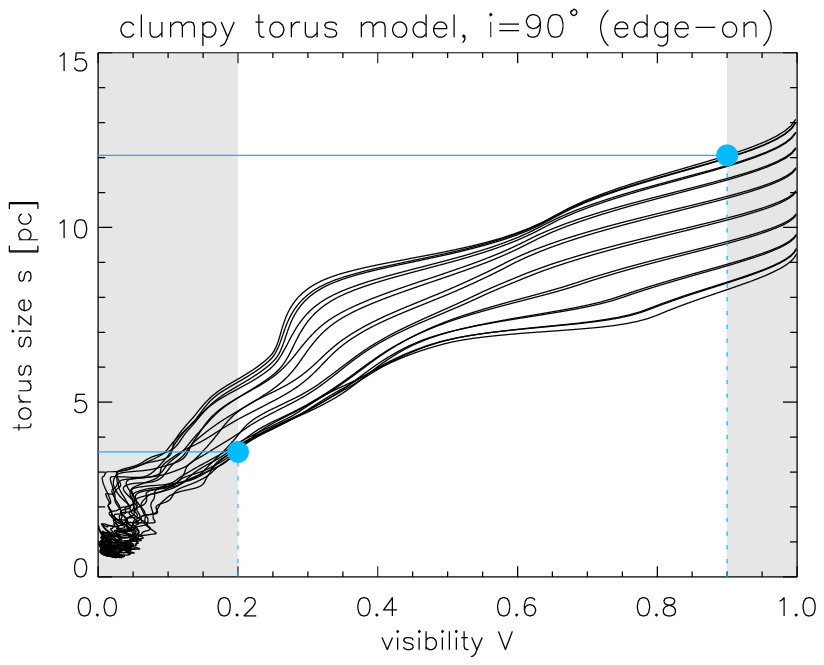}

\caption{Estimated physical torus size as a function of the visibility at $12\,\mathrm{\mu m}$
for the two torus models. \textit{Top row}: full hydrodynamical torus
model (Sect.~\ref{subsec:hydro-models}); \textit{Bottom row}: clumpy
torus model (Sect.~\ref{subsec:rt-models}); \textit{Left panels}:
Seyfert~1 ($i=0^{\circ}$) case; \textit{Right panels}: Seyfert~2
($i=90^{\circ}$) case. Different lines in the plots correspond to
baseline orientations at different position angles. The maxiumum range
of sizes defined by the interval $0.2<V<0.9$ is marked by the blue
lines and dots. The excluded visibilities are shaded in grey.}
\label{fig:size-vis}
\end{figure*}

\subsection{Size estimates for the torus models\label{subsec:modelled-sizes}}

To obtain torus sizes that can be directly compared to the size estimates
from the interferometric measurements, we simulate interferometric
observations of the brightness distributions of the two models. To
this end, we calculate the normalised Fourier transform of the model
images at $12\,\mathrm{\mu m}$ and then extract visibilities at different
baseline lengths for different position angles. This implies that
the single-dish flux measured by MIDI using the $8.2\,\mathrm{m}$
UTs corresponds to the flux integrated over the entire torus model.
This agrees with the emission from the AGN heated cores of almost
all AGN appearing essentially unresolved in the mid-infrared on $8\,\mathrm{m}$
class telescopes \citep[e.g.][]{2009Horst}.

We derive sizes for all model visibilities with $0.2<V<0.9$ assuming
a Gaussian brightness distribution by employing Eq.~\ref{eq:gauss-fwhm},
consistent with the procedure used for the measured visibilities obtained
with MIDI. Since we are interested in the intrinsic uncertainties
in the size estimate due to the Gaussian approximation of the torus
brightness distribution, we do not simulate errors in the visibilities. 

The distance $D$ at which the modelled tori reside is irrelevant
because the spatial frequency in the Fourier domain scales inversely
to the angular size. To observe the same visibilities for a more distant
(i.e. smaller) torus of the same physical size, simply longer baseline
lengths have to be used since $s=D\cdot\mathit{FWHM}\propto D/\mathit{BL}$.
In practice, only a subrange of baseline lengths can be observed with
the VLTI (or any real interferometer), so that only a part or none
(e.g. if the source is unresolved even for the longest available baselines)
of the $0.2<V<0.9$ visibility range can be covered by observations.
By determining the full range of size estimates from the models, we
thus constrain the maximum possible range in observable size estimates.

Fig.~\ref{fig:size-vis} shows the sizes as a function of the visibility
for the two torus models. Each line corresponds to a certain position
angle of the simulated observations. The position angle was changed
in steps of $10^{\circ}$. Along every line, the projected baseline
length changes, leading to different visibilities and hence torus
sizes. As a Gaussian brightness distribution is only an approximation
of the brightness distribution of the respective torus model, the
size estimates calculated according to Eq.~\ref{eq:gauss-fwhm} yield
a range of sizes. If the brightness distribution were a Gaussian,
the calculation would -- of course -- give the same size for all visibilities.

For visibilities $V>0.2$, the size estimates increase monotonically
for growing visibilities. This means that the brightness distributions
of the models have a shallower decrease than a Gaussian brightness
distribution. This is especially true for the clumpy torus model.
Because the clumps of the torus are distributed with a radial distribution
following a power law with an exponent $a=-1$, the brightness distribution
is also similar to a power law, which cannot be closely represented
by a Gaussian brightness distribution.

When the torus is observed face-on (i.e. in the Seyfert~1 case, left
panels of Fig.~\ref{fig:size-vis}), the differences between the
size estimates for different visibilities is dominated by the dependence
on the baseline length, that is on the deviation of the model from
a Gaussian brightness distribution. The total uncertainty in the size
estimate is on the order of $40\ \%$ for the hydrodynamical model
and up to a factor of four for the clumpy torus model. For a torus
seen edge-on (Seyfert~2 case, right panels of Fig.~\ref{fig:size-vis}),
the individual curves show a smaller relative increase in the size
estimate as the visibility increases, especially for the hydrodynamical
model. For this model, this means that the brightness distribution
of the source is actually quite close to a Gaussian brightness distribution.
For small visibilities, $V<0.2$, no unambiguous size can be estimated
from the visibilities, the reason being that these visibilities are
strongly affected by small-scale structures, such as the inner rim
of the torus or the clumps in the dust distribution. These structures
lead to oscillations in the visibility function, i.e. second and higher
order lobes, so that the simplifying approximation by a Gaussian brightness
distribution is no longer applicable%
\footnote{The value of the lower limit to the visibility for an unambiguous
size estimate depends on the strength of the substructure in the torus
brightness distribution. One has to ensure that the visibilities used
for the size estimate are those from the first lobe of the visibility
function. For the torus models analysed here, this is the case for
$V>0.2$, which is also consistent with the limit used in \citet{2009Tristram}.%
}.

In the Seyfert~1 case, no significant dependence of the size estimate
on the position angle of the measurements is present. Because the
tori are seen face-on, they produce brightness distributions that
are roughly circular symmetric (c.f. left column in Fig.~\ref{fig:model-images}).
Hence, the visibilities depend only insignificantly on the position
angle. In the Seyfert~2 case, however, the position angle of the
measurements has a large influence on the size estimate (c.f. the
array of curves in the right panels of Fig.~\ref{fig:size-vis}).
The reason is the significant elongation of the brightness distribution
of a torus seen edge-on. In total, this leads to a difference in the
size estimates in the interval $0.2<V<0.9$ of up to a factor of three.
For the edge-on case of the hydrodynamical model, most of this difference
is due to the intrinsic elongation of the source and not to the approximation
of the brightness distribution by a Gaussian profile. With a sufficient
number of interferometric measurements at different position angles
(three at least), the elongation and orientation of the emission can
be determined. A more accurate, deprojected size estimate for the
torus can then be derived. In the case of the hydrodynamical model,
the uncertainty in the size caused by assuming a Gaussian brightness
distribution can then be reduced to a factor of $1.3$, relative to
the factor of about two for an unknown orientation. Owing to the power-law
dependence of the clumpy torus model, the effect is less pronounced
in this model and the uncertainty is only reduced from a factor of
$\sim3.4$ to $\sim2.5$. In both cases, this uncertainty is lower
than in the Seyfert~1 case. Here, we are primarily interested in
determining which constraints on the dust distributions can be derived
from the simple size estimates alone, without constraining the geometry
of the brightness distribution any further through more complex model
fitting. We thus consider the differences in the size estimates for
different position angles as an uncertainty. The implications of a
clearer knowledge of the geometry are discussed in Sect.~\ref{sub:size-position-angle}.
\begin{figure*}
\centering

\includegraphics[bb=127bp 297bp 466bp 552bp]{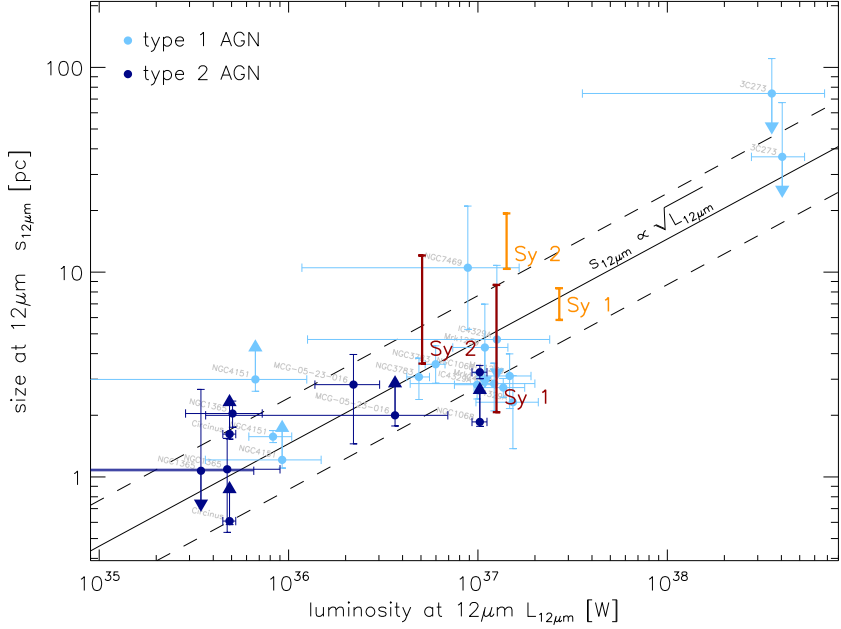}

\caption{Size of the mid-infrared emitter as a function of its monochromatic
luminosity in the mid-infrared for type~1 (light blue) and type~2
(dark blue) AGN. Upper and lower limits on the size estimates are
marked by arrows. The fitted relation $s=p\cdot L^{0.5}$ is delineated
by the black continuous line, the scatter in the relation by the dashed
lines. The ranges of sizes estimated from the hydrodynamical torus
model are plotted in orange, those from the clumpy torus model in
dark red.\label{fig:size-mirsizlum}}

\end{figure*}

The model images used for the calculation represent only one realisation
of the distribution of clumps or filaments. Obtaining size estimates
at different position angles, however, corresponds to probing different
realisations of the fine structure. As can be seen from Fig.~\ref{fig:size-vis},
the fine structure only becomes relevant for the size estimation at
small visibilities, $V<0.2$, that is, when individual parts of the
substructure are being resolved. This means that the size estimates
are independent of any individual realisations of the small-scale
structure in the torus. 

We conclude that using Eq.~\ref{eq:gauss-fwhm} is a viable way of
estimating the size of the brightness distribution from the visibility.
In doing so, the uncertainty in the estimate is on the order of a
factor of three for a torus seen edge-on, with a significant part
of the uncertainty coming from the elongation of the brightness distribution.
The error in the size estimate in the Seyfert~1 case is $\lesssim40\ \%$
for the hydrodynamical model and up to a factor of four for the clumpy
torus model. In this case, the error is dominated by the simplifying
assumption of a Gaussian brightness distribution. For the following,
the range of sizes derived for $0.2<V<0.9$ from the models is considered
as a measure of the expected accuracy of the sizes estimated.

\section{Results and discussion\label{sec:discussion}}

\subsection{Size as a function of the luminosity in the mid-infrared\label{sub:size-mirsizlum}}

The sizes of the warm dust distributions as a function of the luminosity
in the mid-infrared are shown in Fig.~\ref{fig:size-mirsizlum}.
As already found by \citet{2009Tristram}, the sizes are consistent
with $s=p\cdot L^{0.5}$ or $\log\left(s/\mathrm{pc}\right)=q+0.5\cdot\log\left(L/\mathrm{W}\right)$.
The larger number of size estimates allows us to redetermine the proportionality
factor of this relation. We use the median and the standard deviation
of $s\cdot L^{-0.5}$ for the individual measurements to obtain an
estimate for $p$ of $p=(1.45\pm0.20)\cdot10^{-18}\,\mathrm{pc}\,\mathrm{W}^{-0.5}$.
Accordingly, $q=-17.84\pm0.14$. When using, instead of the median,
the mean of all measurements that are not limits, we obtain $p=1.49\cdot10^{-18}\,\mathrm{pc}\,\mathrm{W}^{-0.5}$.
In \citet{2009Tristram}, a proportionality factor of $p=(1.8\pm0.3)\cdot10^{-18}\,\mathrm{pc}\,\mathrm{W}^{-0.5}$
was found. The slightly smaller value now, is mainly due to the additional
measurements for \object{Mrk~1239} and \object{IC~4329A}, which
yield comparatively small size estimates for their respective dust
distributions. That is these two galaxies seem to possess relatively
compact tori in comparison to the other galaxies studied, in particular
in comparison to \object{NGC~4151} and \object{NGC~7469}. In
\citet{2009Tristram}, it was shown that the proportionality factor
was consistent with a more or less optically thick dust distribution
with an average temperature of $300\,\mathrm{K}$. Now a slightly
warmer dust distribution with $T\sim330\,\mathrm{K}$ is consistent
with the measurements. 

Because the sizes are plotted as a function of the luminosity at the
same wavelength, the figure essentially represents a measure for the
\textit{compactness} of the brightness distribution: any size deviating
from the relation implies that the emission is either more compact
(size below the relation) or more extended (size above the relation)
than the average dust distribution.

The ranges of sizes predicted by the models are shown by the thick
bars in Fig.~\ref{fig:size-mirsizlum}. The total fluxes of the models
in the mid-infrared were determined simply by integrating over the
entire modelled brightness distributions. The sizes from the models
are consistent with the observed size estimates, i.e. the torus models
at hand roughly produce the correct torus sizes for a given luminosity.
We nevertheless assert that the emission of the hydrodynamical model
appears to be slightly more extended than expected for the average
torus size according to the measured relation. The torus models can
however not be used to check whether the relation is correct in itself.
It is explicitly assumed that the models scale with $L^{0.5}$, that
is, that they follow the size--luminosity relation in this simple
form. Owing to the large amounts of computing time required, the hydrodynamical
model has been only calculated for a single luminosity. In addition
the clumpy torus model by construction scales as a function of the
inner radius of the torus, $R_{\mathrm{sub}}$ (c.f. Sect.~\ref{subsec:rt-models}).
The current measurements are consistent with this assumption and there
is no indication that the relative sizes of the nuclear dust distributions
change for increasing luminosity other than $\propto L^{0.5}$ as
in the case of some smooth dust distributions or as might occur in
the receding torus paradigm \citep{1991Lawrence}.

Both models show a clear and significant offset in the size estimates
between the Seyfert~1 and the Seyfert~2 cases. The Seyfert~2 case
is shifted towards both a lower luminosity and a greater size with
respect to the same torus seen face-on. Quantitatively speaking, a
torus seen edge-on produces only about half  of the mid-infrared
emission from the same torus seen face-on, even though it is at the
same time roughly twice as extended. The discrepancy is about the
same for both torus models, hence we consider it to be genuine for
all similar toroidal dust distributions. The significance of the discrepancy,
however, is stronger for the hydrodynamical model, because for this
model the uncertainties induced by the assumption of a Gaussian brightness
distribution are smaller. This difference in the apparent mid-infrared
luminosity of the two orientations of the torus has to be taken into
account when evaluating the completeness of mid-infrared selected
AGN samples and the statistics of type 1 and type 2 sources in such
a sample.

For the same mid-infrared luminosity, a Seyfert~1 torus thus appears
about 2.5 times smaller than a Seyfert~2 torus. Its brightness distribution
appears more compact, because it is dominated by the bright emission
from the hot dust at the inner rim of the torus, i.e. it has a stronger
flux concentration towards the centre of the brightness distribution
.This is clearly visible when comparing the images in the left (torus
face-on) and right columns (torus edge-on) of Fig.~\ref{fig:model-images}.
Because this difference in size is significant in the sense that it
is on the order of or larger than the uncertainty in the size estimates,
this should lead to a separation of Seyfert~2 and Seyfert~1 tori
into two distinct loci on the size--luminosity relation. More realistically,
there is of course a continuous distribution of objects between the
face-on and edge-on extremes considered here. That is, there should
be a gradient with the object type in the distribution, perpendicular
to the direction of the relation. However, the scatter in the currently
observed size estimates show that this assumption does not hold in
such a simple way. The apparent differences in individual objects
are much larger than those of the two classes of objects. With the
current data it is thus impossible to ascertain a difference in the
compactness of the type~1 and type~2 tori. With smaller errors and
a larger sample of sources, the statistics may be improved and a distinction
might become possible.%
\begin{figure*}
\centering

\includegraphics[bb=127bp 297bp 466bp 552bp]{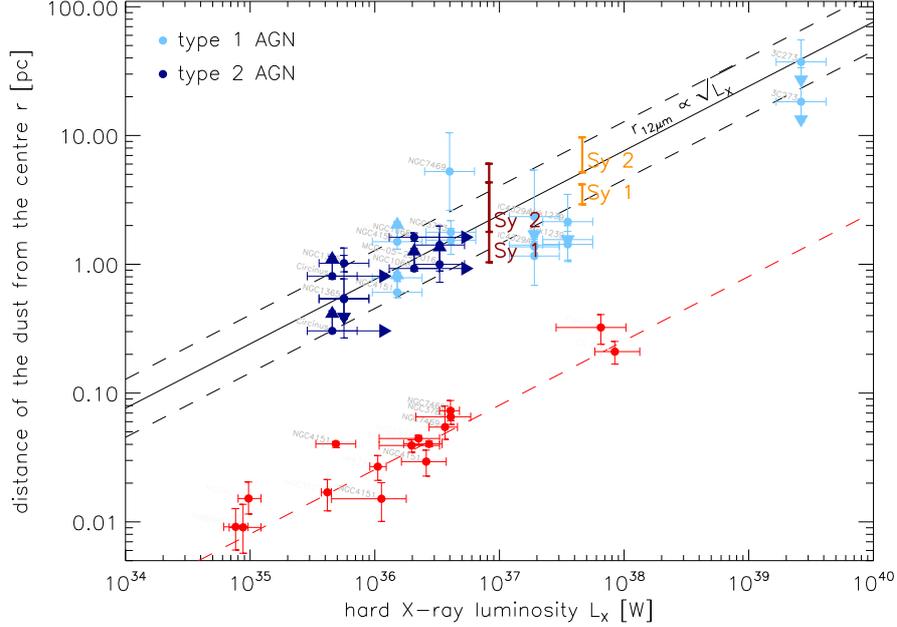}

\caption{Size of the mid-infrared emitter as a function of the X-ray luminosity,
taken as a proxy for the intrinsic AGN luminosity. Colours and symbols
are as in Fig.~\ref{fig:size-mirsizlum}. The red dashed line shows
the inner radius of the torus derived from reverberation measurements
and near-infrared interferometry. The individual measurements of near-infrared
radii are plotted as red points and were taken from \citet{2006Suganuma}.
For these, only the sources common to our sample are labelled.\label{fig:size-bolsizlum}}

\end{figure*}

The strong differences in the accuracy of size estimates derived by
employing a Gaussian approximation for the two different torus models
are caused by their different radial brightness distributions. The
radial brightness profile of the hydrodynamical model is relatively
close to that of a Gaussian profile (especially for the Seyfert~2
case), while that of the clumpy torus model follows a power law. The
scatter in the size estimates for a single object from different visibility
measurements agrees with the uncertainties derived from the models.
Furthermore, lower visibilities yield smaller size estimates. This
agrees well with the monotonically increasing sizes for increasing
model visibilities found in Sect.~\ref{subsec:modelled-sizes} and
indicates that -- unsurprisingly -- the true brightness distributions
deviate from our Gaussian assumption. With more accurate size estimates
for different visibility measurements, it will hence be possible to
distinguish between different radial brightness profiles: a larger
difference between the size estimates at different visibilities will
indicate that the radial brightness distribution has power-law dependence,
as in the case of the clumpy torus model. A weaker dependence would
imply a more Gaussian-like distribution, similar to that of the hydrodynamical
torus model. This result complement methods directly targeted at the
investigation of the radial brightness profile of AGN tori, as for
example carried out by \citet{2009Kishimoto1}. For this, it will,
however, be necessary to significantly reduce the error bars in the
individual, measured size estimates.

\subsection{Size as a function of the estimated intrinsic luminosity\label{sub:size-bolsizlum}}

Instead of plotting the size estimates as a function of the luminosity
in the mid-infrared, $L_{12\mathrm{\mu m}}$, we can also plot the
half-size, that is, the mean distance of the dust from the centre
$r=0.5\cdot s$, as a function of the intrinsic luminosity of the
AGN, using the X-ray luminosities as a proxy. This is shown in Fig.~\ref{fig:size-bolsizlum}
and can be considered to be the {}``true'' size--luminosity relation
and not only a measure of the compactness of the emission region in
the infrared.

The measurements are again consistent with a relation where $r\sim L^{0.5}$
and we obtain $\tilde{p}=(0.76\pm0.11)\cdot10^{-18}\,\mathrm{pc}\,\mathrm{W}^{-0.5}$
for the corresponding proportionality constant. This is not surprising
considering the good correlation between X-ray and mid-infrared luminosities
(see Sect.~\ref{sec:introduction}). The mid-infrared luminosity
itself appears to be a good estimator of the intrinsic AGN luminosity
\citep[e.g. ][]{2009Gandhi}, although we have shown in the previous
section that there is a factor of two difference in the mid-infrared
luminosity of a torus seen face- or edge-on. There might be a trend
for more luminous sources to have smaller-sized mid-infrared emitters
than expected from the size--luminosity relation when considering
the limits on the sizes from type 1 sources with higher luminosities,
that is \object{IC~4329A}, \object{Mrk~1239}, and \object{3C~273}.
However, this deviation from the relation is insignificant compared
to the accuracy of the current data. We note that \object{3C~273}
also has blazar-like properties with the possibility of boosting or
excess non-thermal X-ray emission, which may lead to an overestimate
of its intrinsic luminosity used here.

The size ranges derived for the two torus models are consistent with
the measurements. To plot these models, we simply use $L_{X}=L_{\mathrm{AD}}$,
where $L_{\mathrm{AD}}$ is the total bolometric luminosity of the
heating source used in the models, i.e., the integrated spectrum of
the accretion disk. Hence, this confirms that the X-ray fluxes we
use are closely related to the intrinsic luminosities of the AGN,
assuming that the models are of course correct. Because the amount
of unprocessed energy is now relevant, both the Seyfert~1 and Seyfert~2
cases are located at the same luminosity for both models (leading
to overlap of the two bars indicating the size ranges for the clumpy
torus model).

We also indicate -- with the red dashed line -- the inner radius of
the dust distributions as determined from reverberation measurements
in the $K$ band of several type 1 AGN \citep{2006Suganuma}. To be
able to display that relation in the same plot, we have to correct
their $V$ band luminosities to our hard X-ray luminosities. We adopt
an empirical value of $L_{14-195\,\mathrm{keV}}/L_{V}=3$, so that
the three objects present in both samples (\object{NGC~7469}, \object{NGC~3783},
and \object{NGC~4151}) have about the same luminosity. The correction
factor also agrees reasonably well with values that can be derived
by combining the bolometric correction factors given in \citet{1994Elvis},
\citet{2004Marconi}, and \citet{2009Rigby}. Furthermore, this correction
factor is also consistent with size measurements made of the inner
rim of the torus using $K$ band interferometry from \citet{2011Kishimoto},
when taking into account their correction of $L_{\mathrm{UV}}/L_{V}=6$.
The $K$ band emission is thought to originate from dust at or close
to the sublimation temperature, that is, from the inner rim of the
torus.

We can thus compare the size--luminosity relations for the inner rim
($K$ band) as well as for the body of the torus (mid-infrared) to
investigate the structure of the torus. We find that the dust responsible
for the mid-infrared emission is located about 30 times further outside
than the dust at the sublimation temperature: the warm dust is located
at $0.5\cdot s=r_{\mathrm{body}}\sim30\, r_{\mathrm{in}}$ from the
centre, where $r_{\mathrm{in}}$ is the inner rim of the torus probed
by near-infrared reverberation and interferometry. Our value is similar
to the value of $25\, r_{\mathrm{in}}$ shown in Fig.~1 in \citet{2009Kishimoto1}.
It should be noted that the sizes referred to here are the apparent
sizes of the dust distributions in the infrared, that is, the sizes
of the hot and the warm dust. There may well be significant amounts
of cooler dust at larger radii, which are invisible in the near- and
mid-infrared.

The determined ratio of the size in the mid-infrared to the near-infrared
is an additional constraint of AGN torus models. However, it cannot
directly be used to pin down a certain torus characteristic, as it
depends on many physical parameters, such as the radial dust density
distribution, the global geometrical properties, or the clumpiness
of the torus. Systematic parameter studies with radiative transfer
models of AGN tori are needed to assess this problem and maybe break
the aforementioned degeneracy.

\subsection{Size estimates depending on the position angle\label{sub:size-position-angle}}

The differences in size caused by the elongation of the brightness
distribution for an edge-on torus were considered as an additional
source of uncertainty in the size estimates. By this means, the analysis
of the interferometric data was kept very simple. As noted in Sect.~\ref{subsec:modelled-sizes},
the uncertainties in the size estimates can be reduced if measurements
at different position angles are obtained. Measurements for at least
three different position angles and with similar baseline lengths
are needed to unambiguously determine the elongation (and orientation)
of the source and hence a deprojected size of the dust distribution.
For this, at least an ellipse has to be fitted to the sizes at different
position angles, if a full two-dimensional model fit to the visibility
data is not carried out. The similar baseline lengths for the measurements
at different position angles are important to minimise the influence
of the size dependence on the visibility, that is caused by the Gaussian
approximation. With only two perpendicular measurements, the exact
orientation and elongation will remain unknown, although one can obtain
a mean size of the torus at this baseline length by using the mean
of the two measurements.

The reduced uncertainties in the size estimates will only lead to
a small improvement in the distinction between type~1 and type~2
tori in the size luminosity relation. On the other hand, the presence
of an elongation in the dust distribution alone is a strong discriminator
between face-on and inclined dust tori, as the face-on tori are expected
to show no elongation. A more detailed study of the elongation properties
of the dust distributions will provide strong constraints on torus
models in itself, but is beyond the scope of this paper.

\section{Conclusions\label{sec:conclusions}}

We have carried out a more detailed investigation of the size--luminosity
relation of the warm dust distributions in AGN than the initial study
presented in \citet{2009Tristram} by adding new measurements and
comparing the relation to predictions from hydrodynamical and radiative
transfer models of dusty tori in AGN. The sizes were estimated directly
from the visibilities $V$ assuming a Gaussian brightness distribution
and ignoring any position angle dependence of the size estimates.
The main goal was to ascertain the degree to which such size estimates
can be used to investigate the properties of the nuclear dust distributions
in AGN. 

By calculating the sizes from modelled images of AGN tori, we found
that this method yields viable size estimates with an uncertainty
of up to a factor of four, as long as $0.2\lesssim V\lesssim0.9$.
The uncertainties in the estimated sizes when using this method is
dominated by the simplifying assumption of a Gaussian brightness distribution
as well as the elongation of the emission region for edge-on tori.
Despite their uncertainties, the sizes derived using this simple method
can be used for the investigation of the torus properties and for
a comparison with model predictions.

The sizes derived from images of two different AGN torus models agree
with those determined from the interferometric measurements. According
to the models, there should, however, be an offset between the sizes
of face-on and edge-on tori: face-on tori should be significantly
more compact at the same luminosity, leading to sizes smaller by a
factor of about 2.5 than for edge-on tori. This offset is currently
not observed but may well be hidden in the current uncertainties and
small number statistics of the sizes measured interferometrically
to date. In the present sample, differences between individual objects
are larger than between the two classes of objects and it remains
an open question of whether this observed large scatter is intrinsic,
i.e. due to different tori, or to the uncertainty in the current measurements.
Measurements of a large sample of AGN with higher accuracies than
the current data will be needed to more tightly constrain the relation.
These measurements will be provided soon by the data from our MIDI
AGN Large Programme, which also includes the determination of accurate
total flux spectra using VISIR spectroscopy.

Finally, we have compared the ratio of the sizes in the mid- to the
near-infrared, which probe the body and the inner rim of the torus,
respectively. The resulting value of approximately 30 can be used
to investigate the torus structure, e.g. its geometry and/or its radial
density distribution.

The mid-infrared size--luminosity relation agrees closely with other
AGN scaling relations, such as those for the torus inner radius, $r_{\mathrm{in}}\propto\sqrt{L}$
\citep[e.g.][]{2006Suganuma,2011Kishimoto}, or the size of the broad
line region, $r_{\mathrm{BLR}}\propto\sqrt{L}$ (e.g. \citealt{1990Netzer}
or \citealt{2009Bentz}), with $r_{\mathrm{BLR}}<r_{\mathrm{in}}$.
This is additional evidence that the AGN phenomenon scales with the
luminosity of the central engine.

\begin{acknowledgements}
We thank the anonymous referee for his careful reading of the manuscript
and his valuable suggestions that lead to a significant improvement
of the paper. We especially thank S. F. Hönig for providing images
of his clumpy torus model that were used to obtain the size estimates
for his model as well as W. Jaffe, the PI of programme 381.B-0240.
We also thank L. Burtscher, M. Kishimoto, K. Meisenheimer, and J.-U.
Pott for clarifying discussions and helpful suggestions. This research
has made use of the SIMBAD database, operated at CDS, Strasbourg,
France.
\end{acknowledgements}
\bibliographystyle{aa}
\bibliography{paper}

\begin{thebibliography}{43}
\expandafter\ifx\csname natexlab\endcsname\relax\def\natexlab#1{#1}\fi

\bibitem[{{Absil} {et~al.}(2004){Absil}, {Bakker}, {Schoeller}, \&
  {Gondoin}}]{2004Absil}
{Absil}, O., {Bakker}, E.~J., {Schoeller}, M., \& {Gondoin}, P.~A. 2004, in
  Society of Photo-Optical Instrumentation Engineers (SPIE) Conference Series,
  Vol. 5491, Society of Photo-Optical Instrumentation Engineers (SPIE)
  Conference Series, ed. {W.~A.~Traub}, 1320

\bibitem[{{Antonucci}(1993)}]{1993Antonucci}
{Antonucci}, R. 1993, \araa, 31, 473

\bibitem[{{Beckert} {et~al.}(2008){Beckert}, {Driebe}, {H{\"o}nig}, \&
  {Weigelt}}]{2008Beckert}
{Beckert}, T., {Driebe}, T., {H{\"o}nig}, S.~F., \& {Weigelt}, G. 2008, \aap,
  486, L17

\bibitem[{{Bentz} {et~al.}(2009){Bentz}, {Peterson}, {Netzer}, {Pogge}, \&
  {Vestergaard}}]{2009Bentz}
{Bentz}, M.~C., {Peterson}, B.~M., {Netzer}, H., {Pogge}, R.~W., \&
  {Vestergaard}, M. 2009, \apj, 697, 160

\bibitem[{{Burtscher} {et~al.}(2009){Burtscher}, {Jaffe}, {Raban},
  {Meisenheimer}, {Tristram}, \& {R{\"o}ttgering}}]{2009Burtscher}
{Burtscher}, L., {Jaffe}, W., {Raban}, D., {et~al.} 2009, \apjl, 705, L53

\bibitem[{{Burtscher} {et~al.}(2010){Burtscher}, {Meisenheimer}, {Jaffe},
  {Tristram}, \& {R{\"o}ttgering}}]{2010Burtscher}
{Burtscher}, L., {Meisenheimer}, K., {Jaffe}, W., {Tristram}, K.~R.~W., \&
  {R{\"o}ttgering}, H.~J.~A. 2010, \pasa, 27, 490

\bibitem[{{Cusumano} {et~al.}(2010){Cusumano}, {La Parola}, {Segreto},
  {Ferrigno}, {Maselli}, {Sbarufatti}, {Romano}, {Chincarini}, {Giommi},
  {Masetti}, {Moretti}, {Parisi}, \& {Tagliaferri}}]{2010Cusumano}
{Cusumano}, G., {La Parola}, V., {Segreto}, A., {et~al.} 2010, \aap, 524, A64

\bibitem[{{Elvis} {et~al.}(1994){Elvis}, {Wilkes}, {McDowell}, {Green},
  {Bechtold}, {Willner}, {Oey}, {Polomski}, \& {Cutri}}]{1994Elvis}
{Elvis}, M., {Wilkes}, B.~J., {McDowell}, J.~C., {et~al.} 1994, \apjs, 95, 1

\bibitem[{{Gandhi} {et~al.}(2009){Gandhi}, {Horst}, {Smette}, {H{\"o}nig},
  {Comastri}, {Gilli}, {Vignali}, \& {Duschl}}]{2009Gandhi}
{Gandhi}, P., {Horst}, H., {Smette}, A., {et~al.} 2009, \aap, 502, 457

\bibitem[{{Grupe} {et~al.}(2004){Grupe}, {Mathur}, \& {Komossa}}]{2004Grupe}
{Grupe}, D., {Mathur}, S., \& {Komossa}, S. 2004, \aj, 127, 3161

\bibitem[{{H{\"o}nig} {et~al.}(2006){H{\"o}nig}, {Beckert}, {Ohnaka}, \&
  {Weigelt}}]{2006Hoenig}
{H{\"o}nig}, S.~F., {Beckert}, T., {Ohnaka}, K., \& {Weigelt}, G. 2006, \aap,
  452, 459

\bibitem[{{H{\"o}nig} \& {Kishimoto}(2010)}]{2010Hoenig2}
{H{\"o}nig}, S.~F. \& {Kishimoto}, M. 2010, \aap, 523, A27

\bibitem[{{H{\"o}nig} {et~al.}(2010){H{\"o}nig}, {Kishimoto}, {Gandhi},
  {Smette}, {Asmus}, {Duschl}, {Polletta}, \& {Weigelt}}]{2010Hoenig1}
{H{\"o}nig}, S.~F., {Kishimoto}, M., {Gandhi}, P., {et~al.} 2010, \aap, 515,
  A23

\bibitem[{{Horst} {et~al.}(2009){Horst}, {Duschl}, {Gandhi}, \&
  {Smette}}]{2009Horst}
{Horst}, H., {Duschl}, W.~J., {Gandhi}, P., \& {Smette}, A. 2009, \aap, 495,
  137

\bibitem[{{Horst} {et~al.}(2008){Horst}, {Gandhi}, {Smette}, \&
  {Duschl}}]{2008Horst1}
{Horst}, H., {Gandhi}, P., {Smette}, A., \& {Duschl}, W.~J. 2008, \aap, 479,
  389

\bibitem[{{Jaffe} {et~al.}(2004){Jaffe}, {Meisenheimer}, {R{\"o}ttgering},
  {Leinert}, {Richichi}, {Chesneau}, {Fraix-Burnet}, {Glazenborg-Kluttig},
  {Granato}, {Graser}, {Heijligers}, {K{\"o}hler}, {Malbet}, {Miley},
  {Paresce}, {Pel}, {Perrin}, {Przygodda}, {Schoeller}, {Sol}, {Waters},
  {Weigelt}, {Woillez}, \& {de Zeeuw}}]{2004Jaffe1}
{Jaffe}, W., {Meisenheimer}, K., {R{\"o}ttgering}, H.~J.~A., {et~al.} 2004,
  \nat, 429, 47

\bibitem[{{Kishimoto} {et~al.}(2011){Kishimoto}, {H{\"o}nig}, {Antonucci},
  {Barvainis}, {Kotani}, {Tristram}, {Weigelt}, \& {Levin}}]{2011Kishimoto}
{Kishimoto}, M., {H{\"o}nig}, S.~F., {Antonucci}, R., {et~al.} 2011, \aap

\bibitem[{{Kishimoto} {et~al.}({2009b}){Kishimoto}, {H{\"o}nig}, {Antonucci},
  {Kotani}, {Barvainis}, {Tristram}, \& {Weigelt}}]{2009Kishimoto2}
{Kishimoto}, M., {H{\"o}nig}, S.~F., {Antonucci}, R., {et~al.} {2009b}, \aap,
  507, L57

\bibitem[{{Kishimoto} {et~al.}({2009a}){Kishimoto}, {H{\"o}nig}, {Tristram}, \&
  {Weigelt}}]{2009Kishimoto1}
{Kishimoto}, M., {H{\"o}nig}, S.~F., {Tristram}, K.~R.~W., \& {Weigelt}, G.
  {2009a}, \aap, 493, L57

\bibitem[{{Krolik} \& {Begelman}(1988)}]{1988Krolik}
{Krolik}, J.~H. \& {Begelman}, M.~C. 1988, \apj, 329, 702

\bibitem[{{Lawrence}(1991)}]{1991Lawrence}
{Lawrence}, A. 1991, \mnras, 252, 586

\bibitem[{{Levenson} {et~al.}(2009){Levenson}, {Radomski}, {Packham}, {Mason},
  {Schaefer}, \& {Telesco}}]{2009Levenson}
{Levenson}, N.~A., {Radomski}, J.~T., {Packham}, C., {et~al.} 2009, \apj, 703,
  390

\bibitem[{{Marconi} {et~al.}(2004){Marconi}, {Risaliti}, {Gilli}, {Hunt},
  {Maiolino}, \& {Salvati}}]{2004Marconi}
{Marconi}, A., {Risaliti}, G., {Gilli}, R., {et~al.} 2004, \mnras, 351, 169

\bibitem[{{Matt} {et~al.}(2000){Matt}, {Fabian}, {Guainazzi}, {Iwasawa},
  {Bassani}, \& {Malaguti}}]{2000Matt}
{Matt}, G., {Fabian}, A.~C., {Guainazzi}, M., {et~al.} 2000, \mnras, 318, 173

\bibitem[{{Meisenheimer} {et~al.}(2007){Meisenheimer}, {Tristram}, {Jaffe},
  {Israel}, {Neumayer}, {Raban}, {R{\"o}ttgering}, {Cotton}, {Graser},
  {Henning}, {Leinert}, {Lopez}, {Perrin}, \& {Prieto}}]{2007Meisenheimer}
{Meisenheimer}, K., {Tristram}, K.~R.~W., {Jaffe}, W., {et~al.} 2007, \aap,
  471, 453

\bibitem[{{Nenkova} {et~al.}(2002){Nenkova}, {Ivezi{\'c}}, \&
  {Elitzur}}]{2002Nenkova}
{Nenkova}, M., {Ivezi{\'c}}, {\v Z}., \& {Elitzur}, M. 2002, \apjl, 570, L9

\bibitem[{{Nenkova} {et~al.}(2008{\natexlab{a}}){Nenkova}, {Sirocky},
  {Ivezi{\'c}}, \& {Elitzur}}]{2008Nenkova1}
{Nenkova}, M., {Sirocky}, M.~M., {Ivezi{\'c}}, {\v Z}., \& {Elitzur}, M.
  2008{\natexlab{a}}, \apj, 685, 147

\bibitem[{{Nenkova} {et~al.}(2008{\natexlab{b}}){Nenkova}, {Sirocky},
  {Nikutta}, {Ivezi{\'c}}, \& {Elitzur}}]{2008Nenkova2}
{Nenkova}, M., {Sirocky}, M.~M., {Nikutta}, R., {Ivezi{\'c}}, {\v Z}., \&
  {Elitzur}, M. 2008{\natexlab{b}}, \apj, 685, 160

\bibitem[{{Netzer}(1990)}]{1990Netzer}
{Netzer}, H. 1990, in Active Galactic Nuclei, ed. {R.~D.~Blandford, H.~Netzer,
  L.~Woltjer, T.~J.-L.~Courvoisier, \& M.~Mayor}, 57--160

\bibitem[{{Poncelet} {et~al.}(2006){Poncelet}, {Perrin}, \&
  {Sol}}]{2006Poncelet}
{Poncelet}, A., {Perrin}, G., \& {Sol}, H. 2006, \aap, 450, 483

\bibitem[{{Pott} {et~al.}(2010){Pott}, {Malkan}, {Elitzur}, {Ghez}, {Herbst},
  {Sch{\"o}del}, \& {Woillez}}]{2010Pott}
{Pott}, J., {Malkan}, M.~A., {Elitzur}, M., {et~al.} 2010, \apj, 715, 736

\bibitem[{{Raban} {et~al.}(2009){Raban}, {Jaffe}, {R{\"o}ttgering},
  {Meisenheimer}, \& {Tristram}}]{2009Raban}
{Raban}, D., {Jaffe}, W., {R{\"o}ttgering}, H., {Meisenheimer}, K., \&
  {Tristram}, K.~R.~W. 2009, \mnras, 394, 1325

\bibitem[{{Rigby} {et~al.}(2009){Rigby}, {Diamond-Stanic}, \&
  {Aniano}}]{2009Rigby}
{Rigby}, J.~R., {Diamond-Stanic}, A.~M., \& {Aniano}, G. 2009, \apj, 700, 1878

\bibitem[{{Risaliti} {et~al.}(2005){Risaliti}, {Elvis}, {Fabbiano}, {Baldi}, \&
  {Zezas}}]{2005Risaliti1}
{Risaliti}, G., {Elvis}, M., {Fabbiano}, G., {Baldi}, A., \& {Zezas}, A. 2005,
  \apjl, 623, L93

\bibitem[{{Schartmann} {et~al.}(2010){Schartmann}, {Burkert}, {Krause},
  {Camenzind}, {Meisenheimer}, \& {Davies}}]{2010Schartmann}
{Schartmann}, M., {Burkert}, A., {Krause}, M., {et~al.} 2010, \mnras, 403, 1801

\bibitem[{{Schartmann} {et~al.}(2008){Schartmann}, {Meisenheimer}, {Camenzind},
  {Wolf}, {Tristram}, \& {Henning}}]{2008Schartmann}
{Schartmann}, M., {Meisenheimer}, K., {Camenzind}, M., {et~al.} 2008, \aap,
  482, 67

\bibitem[{{Schartmann} {et~al.}(2009){Schartmann}, {Meisenheimer}, {Klahr},
  {Camenzind}, {Wolf}, \& {Henning}}]{2009Schartmann}
{Schartmann}, M., {Meisenheimer}, K., {Klahr}, H., {et~al.} 2009, \mnras, 393,
  759

\bibitem[{{Suganuma} {et~al.}(2006){Suganuma}, {Yoshii}, {Kobayashi},
  {Minezaki}, {Enya}, {Tomita}, {Aoki}, {Koshida}, \&
  {Peterson}}]{2006Suganuma}
{Suganuma}, M., {Yoshii}, Y., {Kobayashi}, Y., {et~al.} 2006, \apj, 639, 46

\bibitem[{{Swain} {et~al.}(2003){Swain}, {Vasisht}, {Akeson}, {Monnier},
  {Millan-Gabet}, {Serabyn}, {Creech-Eakman}, {van Belle}, {Beletic},
  {Beichman}, {Boden}, {Booth}, {Colavita}, {Gathright}, {Hrynevych},
  {Koresko}, {Le Mignant}, {Ligon}, {Mennesson}, {Neyman}, {Sargent}, {Shao},
  {Thompson}, {Unwin}, \& {Wizinowich}}]{2003Swain}
{Swain}, M., {Vasisht}, G., {Akeson}, R., {et~al.} 2003, \apjl, 596, L163

\bibitem[{{Tristram} {et~al.}(2007){Tristram}, {Meisenheimer}, {Jaffe},
  {Schartmann}, {Rix}, {Leinert}, {Morel}, {Wittkowski}, {R{\"o}ttgering},
  {Perrin}, {Lopez}, {Raban}, {Cotton}, {Graser}, {Paresce}, \&
  {Henning}}]{2007Tristram2}
{Tristram}, K.~R.~W., {Meisenheimer}, K., {Jaffe}, W., {et~al.} 2007, \aap,
  474, 837

\bibitem[{{Tristram} {et~al.}(2009){Tristram}, {Raban}, {Meisenheimer},
  {Jaffe}, {R{\"o}ttgering}, {Burtscher}, {Cotton}, {Graser}, {Henning},
  {Leinert}, {Lopez}, {Morel}, {Perrin}, \& {Wittkowski}}]{2009Tristram}
{Tristram}, K.~R.~W., {Raban}, D., {Meisenheimer}, K., {et~al.} 2009, \aap,
  502, 67

\bibitem[{{Tueller} {et~al.}(2010){Tueller}, {Baumgartner}, {Markwardt},
  {Skinner}, {Mushotzky}, {Ajello}, {Barthelmy}, {Beardmore}, {Brandt},
  {Burrows}, {Chincarini}, {Campana}, {Cummings}, {Cusumano}, {Evans},
  {Fenimore}, {Gehrels}, {Godet}, {Grupe}, {Holland}, {Kennea}, {Krimm},
  {Koss}, {Moretti}, {Mukai}, {Osborne}, {Okajima}, {Pagani}, {Page}, {Palmer},
  {Parsons}, {Schneider}, {Sakamoto}, {Sambruna}, {Sato}, {Stamatikos},
  {Stroh}, {Ukwata}, \& {Winter}}]{2010Tueller}
{Tueller}, J., {Baumgartner}, W.~H., {Markwardt}, C.~B., {et~al.} 2010, \apjs,
  186, 378

\bibitem[{{Wittkowski} {et~al.}(2004){Wittkowski}, {Kervella}, {Arsenault},
  {Paresce}, {Beckert}, \& {Weigelt}}]{2004Wittkowski}
{Wittkowski}, M., {Kervella}, P., {Arsenault}, R., {et~al.} 2004, \aap, 418,
  L39

\end{thebibliography}
\end{document}